\documentclass[fleqn,usenatbib]{mnras}

\usepackage{newtxtext,newtxmath}

\usepackage[T1]{fontenc}

\DeclareRobustCommand{\VAN}[3]{#2}
\let\VANthebibliography\thebibliography
\def\thebibliography{\DeclareRobustCommand{\VAN}[3]{##3}\VANthebibliography}

\usepackage{graphicx}	
\usepackage{amsmath}	

\title[Disc Photoevaporation \& Resonant Giant Planets]{The impact of disc photoevaporation on the long-term evolution of giant planets in mean motion resonances}

\author[Greenfield \& Owen]{
Emmanuel J. Greenfield,$^{1}$\thanks{E-mail: eg319@ic.ac.uk}
James E. Owen,$^{1}$
\\
$^{1}$Astrophysics Group, Department of Physics, Imperial College London, Prince Consort Road, London SW7 2AZ, UK\\
}

\date{Accepted XXX. Received YYY; in original form ZZZ}

\pubyear{2025}

\begin{document}

\newcommand{\rc}[1]{\textcolor{black}{#1}}

\label{firstpage}
\pagerange{\pageref{firstpage}--\pageref{lastpage}}
\maketitle

\begin{abstract}
We investigate the long-term impact of disc photoevaporation on the dynamical stability and evolution of giant planet pairs in mean motion resonances. Using two-dimensional hydrodynamical simulations with FARGO3D, in which we have included mass-loss due to photoevaporation, we explore a parameter space covering disc mass, viscosity, planet mass, and resonance type. We find that strong photoevaporation depletes gas in the common gap between the planets, slowing migration and suppressing planet–disc interactions that typically lead to resonance breaking and eccentricity damping. This stabilising effect is most significant for 3:2 resonances, which are more prone to disruption due to the reduced planet spacing. In contrast, 2:1 resonances are generally more robust but can still be destabilised at high disc mass and moderate-to-strong photoevaporation due to asymmetric torques. Photoevaporation can therefore stabilise resonances that would otherwise break, or conversely disrupt resonances that are natively more stable. Even in cases where photoevaporation does not directly affect resonance stability, it typically results in increased planetary eccentricities. These results highlight the complex, system-dependent nature of resonance evolution, with implications for the final orbital architectures of giant planet systems and their detectability via astrometry from missions such as Gaia.

\end{abstract}

\begin{keywords}
protoplanetary discs -- planet–disc interactions 
\end{keywords}



\section{Introduction}

The current exoplanet population contains a rich and diverse arrangement of planets. In this population, we find many multi-planetary systems, and amongst these there is evidence for many systems to be resonant or near-resonant, for both low \citep[e.g.][]{fabrycky_architecture_2014} and high mass systems \citep[e.g.][]{winn_occurrence_2015}. These observations give invaluable information on the dynamical history of these systems; in particular, they highlight the importance of planet-planet interactions during the formation process. Furthermore, observations suggest that young giant planet systems such as PDS 70 \citep{haffert_two_2019} and HR 8799 \citep{marois_images_2010}, both possibly contain planets in resonances. This result could indicate that resonances are a relatively common outcome of the planet formation process. The dynamical history of exoplanet systems is conversely critical to understand recent observations of planetary properties. For example, different close-in giant planet migration pathways can lead to different obliquities and potentially different observable atmospheric compositions \citep{kirk_bowie-align_2024}. Therefore, understanding the dynamical histories of different exoplanet populations can provide additional context in which to interpret a wide variety of different observations. 

The most extensive data on giant planets that we currently have come from radial velocity (RV) and transit surveys, but future data releases from Roman microlensing \citep{penny_predictions_2019} and Gaia astrometry \citep{casertano_astrometric_1995} will be especially sensitive to detecting cool giant planets \citep{perryman_astrometric_2014, lammers_exoplanet_2025}. In fact, astrometry data from Gaia should be precise enough to give us detailed information about the planets' orbital architectures. In preparation for the upcoming Gaia DR4, it is important to understand the dynamical processes that can affect the orbital evolution of large separation giants. In particular, the protoplanetary disc in which giant planets form has a significant impact on the final system architecture. The core accretion model \citep{pollack_formation_1996,rafikov_atmospheres_2006,bailey_growing_2024} generally points towards the outer disc ($\gtrsim 1$ \&~ $\lesssim 20$~AU) as a prime location for giant planet formation, due to the higher solid mass reservoir that can be accreted. 

There have been attempts to match features in observed discs with potential planets \citep{lodato_newborn_2019,ruzza_dbnets_2024,ruzza_dbnets20_2025}; assuming all these features are related to planets they indicate a higher occurrence of young giant planets at 10-100 AU. Contrasting these results with RV surveys of large orbit giants, e.g., the California Legacy Survey \citep{fulton_california_2021} and the \citet{Mayor2011} survey analysed by \citet{fernandes_hints_2019}, known giant planets have a rising occurrence rate out to the 1-10 AU range, with a tentative decline further out. These results suggest that the early dynamical interactions are possibly controlling the giant planet architectures we observe billions of years later. The properties of the giant planet distribution are not yet fully understood, and gathering more data through RVs is difficult given the long baseline for observations. Gaia, however, will be able to rigorously confirm these trends by gathering data on many more planets and study the hypothesised  prime location for wide-orbit giant planet formation by probing these regions in more depth. Although dynamical evolution during the early stages of system lifetime is important, it is clear that dynamical interactions after disc dispersal are also significant \citep{
frelikh_signatures_2019,schoettler_effect_2024}.

Planet-disc interactions play an important role in driving the orbital evolution of young protoplanets  \citep{lin_tidal_1979,goldreich_excitation_1979}. When the planets reach a certain mass, the spiral shocks that they launch through the disc are strong enough to counter viscous spreading, depleting the co-orbital region and creating a gap \citep{goldreich_disk-satellite_1980,lin_structure_1980,kanagawa_formation_2015}, which significantly changes the planet-disc interaction picture and sets the stage for Type II migration \citep{lin_tidal_1986,kanagawa_radial_2018}. Giant planets are generally thought to belong to this regime and migrate slowly, typically inwards, until the disc is dispersed after a timescale of a few Myr \citep{hernandez_spitzer_2007,ribas_protoplanetary_2015}.
A second giant planet creates an additional gap of its own, thus also indirectly affecting planet-disc interactions of the first planet \citep{baruteau_disk-planets_2013}. In some cases, the planets' converging migration rates will lead the gaps to overlap, generating a ``common gap'', creating another new picture for planetary evolution. In many cases, the different migration rates lead the planets to cross mean motion resonances, where planet-planet interactions can  lead to resonant trapping or repulsion \citep{rometsch_migration_2020,kanagawa_radial_2020}.  

The disc's profile has an impact on the disruption of such resonances, especially in the region closest to the planets, and in the case of multiple planets, the structure of the common gap. Since the disc's profile evolves over its lifetime, a stable resonant configuration at one point in time might be disrupted later on. For example, \citet{liu_dynamical_2017} looked at the case where a pair of resonant super-earths reaches the inner disc cavity, and one planet reached an orbit inside the cavity. They found the modified asymmetric torque profile could then disrupt the resonance. Given that giant planets migrate slower, we must consider the effects that may impact the surface density in the outer disc. Disc dissipation through viscous accretion predicts much longer lifetimes for discs than observed, and the colour-colour evolution of young stars points to inside-out dispersal \citep{ercolano_clearing_2011, koepferl_disc_2013,lovell_alma_2020}. These observations motivated the development of photoevaporation models as an additional mechanism to remove disc mass. \citep{hollenbach_photoevaporation_1994,clarke_dispersal_2001,owen_radiation-hydrodynamic_2010,sellek_photoevaporation_2024}.
High-energy radiation ionises the upper layer of the disc and heats up the gas, giving it enough energy to drive an outflow.
This high-energy radiation can come either from nearby massive stars in the stellar birth cluster \citep{johnstone_photoevaporation_1998,haworth_fried_2018} or from the central protostar itself. The latter mechanism has been explored for single planets \citep{rosotti_interplay_2013,rosotti_long-term_2015}, where the authors ran hydrodynamical simulations and found that the presence of giant planets accelerates inner disc dispersal from photoevaporation. Since in most models the surface density mass loss profile from photoevaporation reaches a maximum at a few AU, the evolution of giant planets is likely strongly influenced at these orbital separations \citep{jennings_comparative_2018, monsch_giant_2021}. 

In the scenario where two giant planets share a common gap, the lower gas density in this gap makes it significantly more sensitive to photoevaporation, especially considering that viscosity cannot efficiently refill the material in the common gap. This sensitivity, in conjunction with the fact that the inner disc can be more easily depleted through photoevaporation in the presence of giant planets, means that the dynamical evolution of multiple giant planets in a photoevaporating disc could be important. All of these effects have not yet been combined in long-term hydrodynamical simulations.  In this study,  we aim to characterise the effects of disc photoevaporation on the evolution and resonant stability of two giant planets in an evolving protoplanetary disc.

\section{Methods}

We make use of the hydrodynamic code FARGO3D \citep{benitez-llambay_fargo3d_2016}, which we have modified to include the impact of photoevaporation on multiple migrating planets. 

\subsection{Disc model and parameter study}

To determine global disc properties we assume a locally isothermal disc spanning 0.1-150 AU (but only simulate the part of it relevant to this study, as discussed in section \ref{Grid}). We assume an initial surface density profile $\Sigma(R)$, following an inverse power law:

\begin{equation}
    \Sigma(R) = \Sigma_0 \left(\frac{R}{1 \,\mathrm{AU}}\right)^{-1},
\end{equation}

where $R$ is the cylindrical radius. We fix $\Sigma_0$ according to the total mass $M_{\text{d}}$ of the unperturbed disc, set to 0.5, 1.5 and 4.5 per cent of the central star mass, taken to be one solar mass. 

The disc's scale height ($H$) increases with radial distance according to the passive disc flaring profile \citep{kenyon_spectral_1987}:

\begin{equation}
    \frac{H}{R} = h_0 \left(\frac{R}{1 \,\mathrm{AU}}\right)^{0.25}.
\end{equation}

$h_0$ is calculated from the requirement of having a disc temperature of 300K at 1AU, and as such, is set in all setups to be $0.02$, which is lower than standard values used in similar studies. We compare our results for $h_0=0.03$ in Appendix \ref{scaleheight3}, finding our conclusions are robust.
Viscosity is included through the $\alpha$ prescription of \citet{shakura_black_1973}, where a constant value of $\alpha = 10^{-3}$ is used throughout, except for one simulation where $\alpha = 10^{-4}$. The full description of the parameter study is presented in Table \ref{tab:param_study}. We also include the \texttt{BM08} flag to compensate for the effects of disc self-gravity \citep{baruteau_type_2008}. 

\subsection{Planets}

The planets are included in the simulations as additional potentials $\Phi_\text{p}$, each of the form:

\begin{equation}
    \Phi_\text{p} = - \frac{Gm_\textrm{p}}{\sqrt{r^2 + s^2}}
    \label{smoothing}
\end{equation}
where $G$ is the gravitational constant, $m_\textrm{p}$ is the mass of the planet and $r$ is the distance to the planet. To avoid singularities, we use the smoothing parameter $s = 0.6 H(R)$. Since a planet also induces an acceleration on the star, we include the indirect term from that interaction.

For the purpose of this study, we need to include giant planets capable of building a deep gap, which allows disc photoevaporation to become locally important. A fiducial mass of 1 $\text{M}_\text{J}$ was chosen for the two planets and the effect of planets of different masses is also investigated as detailed in Table \ref{tab:param_study}. 

To ensure a controlled experiment of photoevaporation's impact, each run was set up in three steps. First, the planets are initialised outside of commensurability, with the outer planet at 15 AU and the inner planet at a period ratio of $P_2/P_1 = 2.2$ for a 2:1 resonance and $1.6$ for a 3:2 resonance. During this step, the planets' evolution does not take disc torques into account, and they steadily build a gap.

To make sure that the gap is deep enough to allow for realistic migration, we track the evolution of the first planetary gap surface density minimum. Modelling this minimum surface density as a decaying exponential, we fit the first few data points. We consider that the final gap depth has been reached after 5 decay lifetimes, and hence planet-disc and planet-planet interactions are only included once the simulation has reached this point.
Once this point is met, the planets are allowed to migrate and can approach resonance. 
Once the resonant argument is found to librate, the simulation enters the third phase, in which the different photoevaporation strengths are applied to the simulation. The runs stop once the inner disc is dispersed, the planets scatter and get ejected, or the resonance is broken (i.e. the resonant angle is found to circulate).

\subsection{Photoevaporation}

FARGO3D is used to solve the fluid mass and momentum conservation equations in 2D, on a polar grid. Disc photoevaporation acts as a sink of surface density, as such, we modify FARGO3D to solve the following continuity equation:
\begin{equation}
    \frac{\partial \Sigma}{\partial t} + \nabla \cdot (\Sigma \boldsymbol{v}) = - \dot{\Sigma}_{\text{PE}}
\end{equation}

where $\boldsymbol{v}$ is the gas velocity and $\dot{\Sigma}_{\text{PE}}$ is the local surface density mass loss rate due to photoevaporation.
Since this study is only concerned with the effects of disc mass loss due to photoevaporation, the exact model chosen matters less than works with focus on global disc evolution (see the discussion in section \ref{PEmodelchoice}). 

\begin{figure}
	\includegraphics[width=\columnwidth]{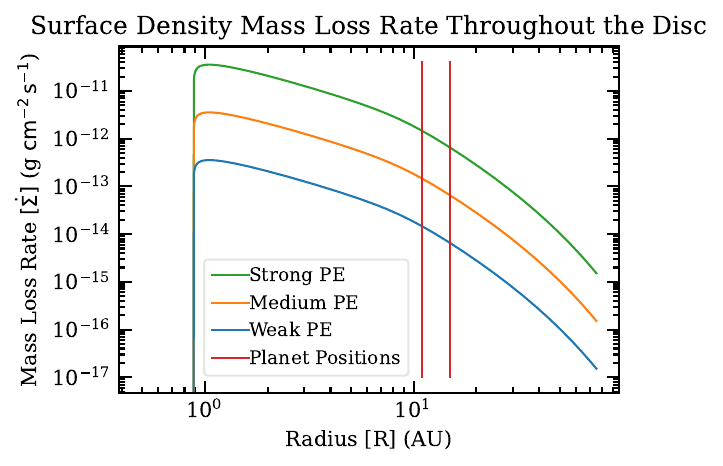}   
    
    \caption{The photoevaporation mass loss profile used in our simulations for each photoevaporation strength, from \citep{alexander_dust_2007}. The initial planet profiles are indicated for the fiducial setup (see section \ref{results}). These profiles are only valid in the diffuse radiation field regime, where photoevaporation only happens from the upper layers of the disc. This model prescribes mass loss at radii $\gtrsim$ 0.1 $R_\text{g}$. Given the disc inner boundary is at 1 AU, this means that the entirety of the simulated disc experiences mass loss. The planets are at about 10 AU, where this surface mass loss rate is roughly an order of magnitude weaker than at its peak in the inner disc, although the typical local mass-loss rate $\sim R^2\dot{\Sigma}_{\text{PE}}$ is larger. }
    \label{fig:wind_density}
\end{figure}

Photoevaporation models use the gravitational radius $R_\text{g}$ as a measure of the radius in the disc beyond which the gas in the upper layers can freely escape from the disc, although mass-loss does occur inside $R_g$ \citep[e.g.][]{Liffman2003}. It is expressed in \citet{hollenbach_photoevaporation_1994} as:
\begin{equation}
    R_g = \frac{GM_*}{c_s^2}\,
\end{equation}
and is 8.84 AU in our setup. 
Due to its simplicity, we have decided to implement the 1D EUV photoevaporation surface density mass loss rate parametrised in \citet{alexander_dust_2007}, where mass loss from photoevaporation is found to occur outside 0.1 $R_\text{g}$, and this model is represented in Fig. \ref{fig:wind_density}.

Photoevaporation was added as an additional source substep, to be executed before the other source substeps. It is implemented as a first-order in time finite difference scheme, removing the appropriate amount of disc material from each cell at each timestep.
We study the effect of three different photoevaporation rates: $\dot{M}_{\text{weak}} =10^{-10} \: \mathrm{M_\odot \,yr^{-1}}$, $\dot{M}_{\text{medium}} =10^{-9} \:\mathrm{M_\odot\,yr^{-1}}$, and $\dot{M}_{\text{strong}} =10^{-8} \:\mathrm{M_\odot\,yr^{-1}}$, where $\dot{M}$ is found by integrating $\dot{\Sigma}_{\text{PE}}$ over the full disc's size (0.1 to 150 AU). The higher end of this range runs the risk of completely depleting the disc density in the gap, so a surface density floor value was implemented in order to prevent negative values. Different values of the floor value were tested, and the optimal floor was chosen to be the highest surface density that negligibly impacted the evolution of the disc and planets. Setting it too low would, however, significantly slow down the simulations. This floor value was chosen as $10^{-4}\: \mathrm{g~cm^{-2}}$.

\subsection{Grid setup and boundaries} \label{Grid}

The grid setup was chosen to encapsulate the full extent of the planets' influence on the disc, spanning 1-75 AU with a logarithmic radial spacing.

Long-term planetary evolution can be sensitive to the inner boundary, especially if it migrates, if the boundary is not considered carefully. On top of the inner boundary there is also a damping zone that affects the planet if it gets too close, regardless of grid resolution. Thus, our approach is motivated by what has been done in previous work to allow for migrating planets in an evolving disc \citep{cimerman_formation_2018}. The damping zone is set to be at a Keplerian orbit with 1.15 times the orbital period of the inner boundary (1 AU), and considering that the furthest Lindblad torque for a planet is around at 2:1 resonance \citep{goldreich_excitation_1979}, we find that the minimal distance for the planet's furthest inner Lindblad torque to be in the damping zone is 1.83 AU. We therefore have decided to take the value of 2.5 AU to calculate the grid resolution. This higher value was chosen to compensate for any other higher order effects, as well as to avoid over-inflating the grid resolution which would result in unnecessarily long runtimes.
At this inner effective limit, we define the resolution of 8 radial cells per scale height, resulting in 1368 radial cells in the base case where $h_0 = 0.02$. We chose to use 1990 azimuthal cells to preserve a square cell shape, thus minimising potential sources of numerical noise. The planets were all retrospectively checked at the end of the simulation, and none have migrated to orbital distances closer than 2.5 AU, meaning that the grid resolution does not affect our results.

Our study aims to investigate the long-term evolution of planets in a disc; therefore, the boundary conditions need to reflect the appropriate evolution in surface density, due to viscous accretion as well as photoevaporation. 
 The damping zones implement the wave damping recipe from \citet{de_val-borro_comparative_2006}, 

\begin{equation}
    \frac{dx}{dt} = \frac{x-x_0}{\tau} R(r)
\end{equation}

where $x$ is the quantity to damp (surface density or gas velocity), $x_0$ is the target to be damped to, $\tau = 0.3 \: \Omega_K^{-1}$ is the damping timescale and $R(r)$ is a parabolic function that increases from 0 to 1 in the damping zone. 

Although the presence of planets greatly alters the radial velocity profile locally, test runs have shown that the damping zones are far enough away from the planet to consider that these effects will not dominate and a constant value of radial velocity, calculated from viscous accretion, is assumed in the damping zones. 
The surface density cannot be treated with the same simplicity, since the effects of accretion and photoevaporation accumulate over long simulation times (greater than the viscous time). Instead, the value of $x_0$ needs to depend on location and time. 
An appropriate way to include this dependence is by calculating the azimuthally averaged surface density for each radial cell within the damping boundary and choose this quantity as the local target. This averaging allows for the cancellation of azimuthal waves, all while allowing for global radial density evolution. 
The azimuthal velocity is not damped at all here, to preserve any deviations from a Keplerian profile, though the ghost cells are set to a pressurised Keplerian azimuthal profile.

A custom set of boundary conditions has also been implemented in the ghost cells to allow the disc to be depleted due to accretion. The value of the surface density in the ghost cells is set to match the mass accretion rates so that $\dot{M}_{\mathrm{active}} = \dot{M}_{\mathrm{ghost}}$. The radial velocity is extrapolated as such: if there is a negative flow, towards the origin, then a symmetric boundary is applied. If the flow is positive, outwards from the origin, there is an antisymmetric boundary condition applied. 

This boundary condition is especially important at the inner boundary to prevent any non-physical outflows into the disc, and at the same time allow for viscous evolution at the boundary. The impact on the outer boundary is less significant as the viscous timescale in those regions is 5000 times longer than the inner boundary.

\begin{figure*}
	\includegraphics[width=\textwidth]{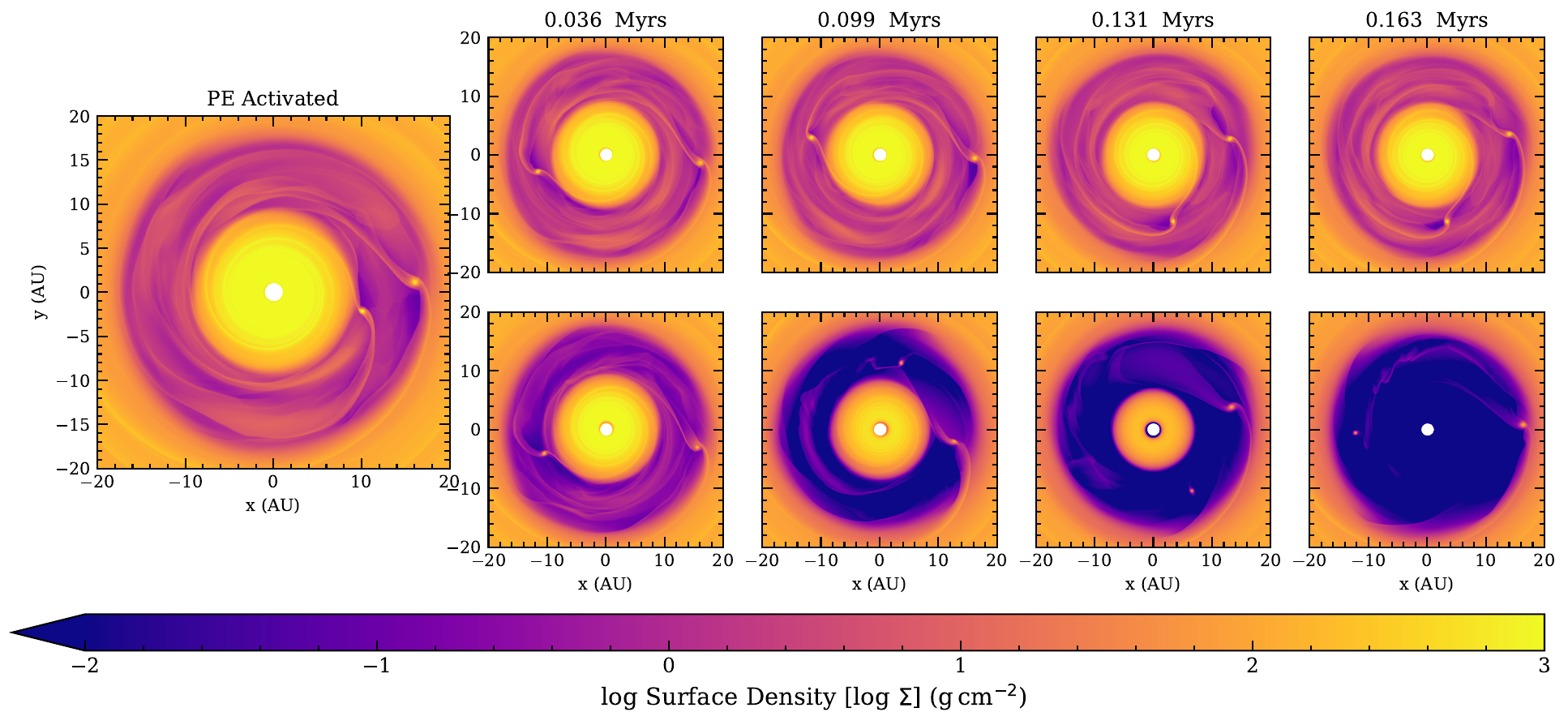}
    \caption{Surface density profiles of the disc at different snapshots, with time shown indicating the time after photoevaporation was activated. The top row shows a disc without photoevaporation and the bottom row shows the results of a disc with $\dot{M}_{\text{PE}} = 10^{-8} \mathrm{M_{\odot}.yr^{-1}}$. The impact of photoevaporation is clear in the common gap at 0.036 Myr, making it deeper. Once the gap is depleted (0.099 Myr onwards), the inner disc is very quickly dispersed since the flow of mass from the outer disc is insufficient to replenish it.}
    \label{fig:2Dsurfdens}
\end{figure*}

\begin{figure}
	\includegraphics[width=\columnwidth]{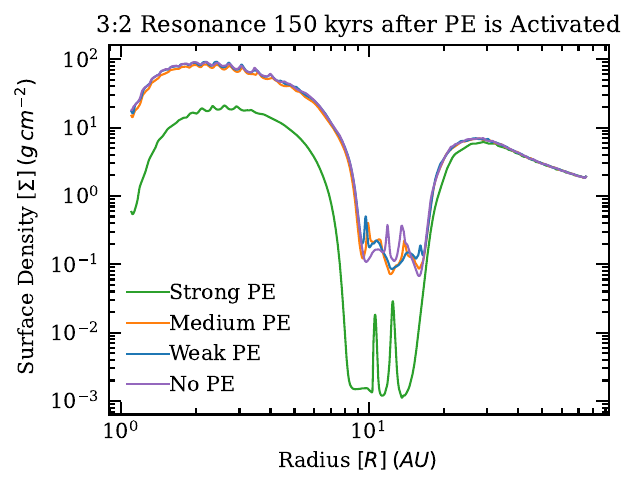}   
    
    \caption{The azimuthally averaged disc surface density profiles at 0.15 Myr after photoevaporation is activated for each photoevaporation strength. The inner disc boundary surface density drops in all cases due to the boundary conditions, but is more strongly depleted (by a factor of about 10) in the strong photoevaporation case. We notice a two order of magnitude drop of the surface density in the common gap only in the case of strong photoevaporation.}
    \label{fig:1Dsurfdens}
\end{figure}

\begin{table}
	\centering
	\caption{This table shows each simulation and what parameters are varied for each one of them.}
	\label{tab:param_study}
	\begin{tabular}{lcccr} 
		\hline
		Name & Resonance & $M_{\text{disc}}$ ($M_*$) & $\text{M}_\text{p}$ ($\text{M}_\text{J}$) & $\alpha$ \\
		\hline
		2t1dm05 & 2:1 & 0.5\% & (1,1) & $10^{-3}$  \\
		2t1dm15 & 2:1 & 1.5\% & (1,1) & $10^{-3}$\\
		2t1dm45 & 2:1 & 4.5\% &  (1,1)& $10^{-3}$\\
            3t2dm05 & 3:2 & 0.5\% & (1,1) & $10^{-3}$\\
		3t2dm15 & 3:2 & 1.5\% & (1,1) & $10^{-3}$\\
		3t2dm45 & 3:2 & 4.5\% &  (1,1)& $10^{-3}$\\
		\hline
        3t2dm15a4 & 3:2 & 1.5\% &  (1,1)& $10^{-4}$\\
         3t2dm15m12 & 3:2 & 1.5\% &  (1,2)& $10^{-3}$\\
         3t2dm15m21 & 3:2 & 1.5\% &  (2,1)& $10^{-3}$\\
         3t2dm15m22 & 3:2 & 1.5\% &  (2,2)& $10^{-3}$\\
         \hline
	\end{tabular}
\end{table}

\section{Results} \label{results}

We first present the results for the fiducial simulation, then we examine the different results across the broader parameter space.

\subsection{Impact on the disc's structure}

\begin{figure}
	\includegraphics[width=\columnwidth]{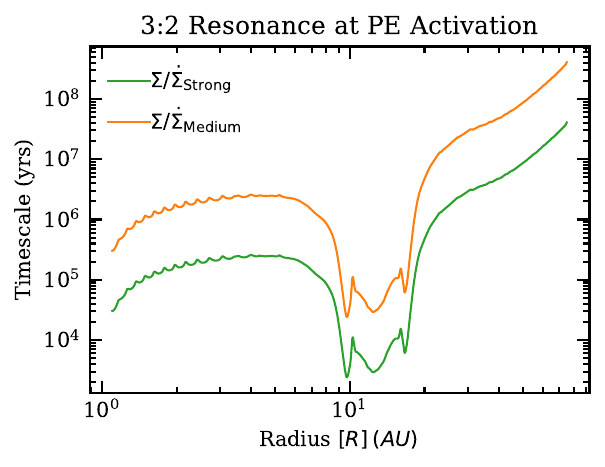}   
    
    \caption{Comparing the local photoevaporative depletion timescales ($\Sigma/\dot{\Sigma}_{\text{PE}}$) for medium and strong photoevaporation. Outside the gap, only the strong photoevaporation timescale is below the viscous time, explaining the features in Fig. \ref{fig:1Dsurfdens}. In the gap region, both strong and medium photoevaporation timescales are below the viscous timescale. Planetary torques contribute to accretion and are not taken into account, explaining why we do not observe any gap deepening in Fig. \ref{fig:1Dsurfdens} for medium photoevaporation. The photoevaporation timescale is an order of magnitude lower for strong photoevaporation, explaining why this is the only setup with a deeper gap in the presence of photoevaporation.}
    \label{fig:timescales}
\end{figure}

To characterise the effects of disc photoevaporation on giant planet resonances, we first look at the fiducial case of planets in a 3:2 resonance, in a disc with $M_{\text{disc}} = 1.5 \% \,M_*$, $\alpha = 10^{-3}$ and $h_0 = 0.02$. We look for differences in the local environment of the planets that impact their dynamical evolution.
The main effect of disc photoevaporation is clearly demonstrated in Fig. \ref{fig:2Dsurfdens}, where the disc's surface density evolution is shown in the case of strong photoevaporation, compared to the case with no photoevaporation. The main difference is  that the common gap between 10 and 17 AU is significantly more depleted and deeper, as shown in the lower panel. 

The gap's depletion is also shown in Fig. \ref{fig:1Dsurfdens}, where the azimuthally averaged disc density evolution is compared for different photoevaporation strengths. This gap depletion appears only to be significant for strong photoevaporation. Although the photoevaporative mass loss profiles from \citet{alexander_dust_2007} peak at 0.9 AU (as seen in Fig. \ref{fig:wind_density}), the photoevaporation depletion timescale $\tau_{\text{PE}}=\Sigma/\dot{\Sigma}_{\text{PE}}$ is the lowest in the common gap between 10 and 20 AU, due to the lower local surface density, as seen in Fig. \ref{fig:timescales}. Although the depletion timescale is lower than the viscous timescale in the gap by at least an order of magnitude, planetary torques contribute to accretion, explaining why there is no impact on the gap structure for medium photoevaporation. For photoevaporation, the depletion timescale is even shorter in the gap overcoming accretion from the planetary torques, causing the gap structure to change.

Fig. \ref{fig:1Dsurfdens} also highlights the depletion of the inner disc due to photoevaporation for strong photoevaporation. This inner disc depletion in the presence of a giant planet due to photoevaporation is similar to what was found by \citet{rosotti_interplay_2013}. Outside the gap, the planetary torques are weaker, so the viscous picture is more accurate. In other words since the accretion rate into and through the gap is higher than the integrated loss to photoevaporation in the medium photoevaporation case, the inner disc can be resupplied. However, in the case of strong photoevaporation, the accretion rate supplying the inner disc is suppressed sufficiently such that it cannot be resupplied, leading to inner disc clearing.

Another consequence of photoevaporation on planetary evolution is that the faster disc depletion leaves less time for the planets to interact with the disc, thereby potentially reducing the time during which they migrate and their migration speed (as discussed in Section \ref{migration_speed}). The gap being depleted faster also means that there is less material in the gap for the planets to interact with, hence reducing the potential planet-disc interactions that can lead to resonance breaking (see Section \ref{resevo}).

\subsection{Resonance evolution} \label{resevo}

\begin{figure}
	\includegraphics[width=\columnwidth]{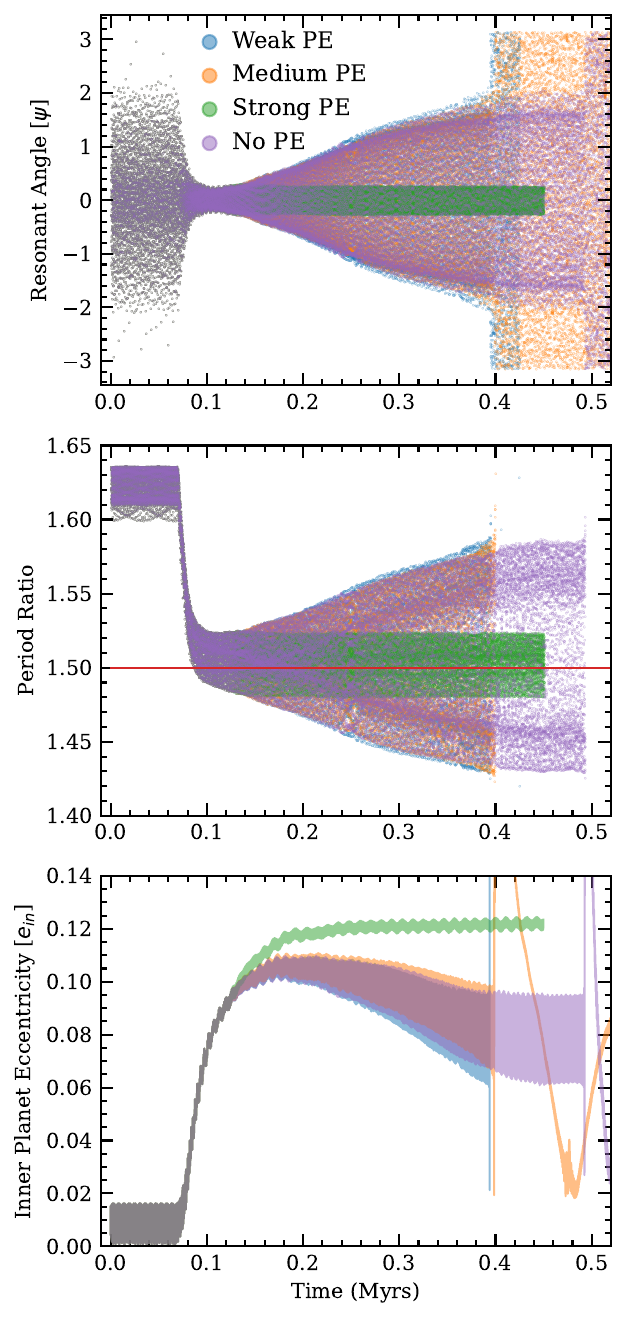}
    \caption{Evolution of the resonant angle (top), planet period ratio (middle) and inner planet eccentricity in the nominal case (bottom). The first phase of the simulation has the planets clear out a gap without any other interactions. During this phase, the pattern in the resonant angle plot is due to the output sampling frequency and has no physical significance. The second phase starts when the planets migrate until they reach resonance (0.1 Myr). At this point, photoevaporation is activated and the simulation enters phase 3. With weak or no photoevaporation, the resonant angle oscillations become larger and larger: a sign of resonance breaking. The period ratio also takes values further and further away from exact commensurability and the eccentricity slowly drops. Strong photoevaporation limits all of these effects, with the eccentricity climbing and the resonance remaining stable until the inner disc dissipates.}
    \label{fig:nominalres}
\end{figure}

In order to determine whether or not the two planets are in resonance, we use the resonant argument:
\begin{equation}
    \psi = (p+q)\lambda_{2} - p \lambda_1 - q\varpi_1,
\end{equation}
where $p$ and $q$ are integers representing a $p+q:q$ resonance, $\lambda_i$ is the longitude of planet $i$ ($i=1$ and $i=2$ for the inner and outer planets respectively), and $\varpi_1$ is the longitude of the pericentre of the inner planet, \citep{murray_solar_1999}. The resonant argument is essentially a measure of the longitude at which a planetary conjunction occurs. A stable first-order resonance will have this angle librate around either the pericentre or apocentre, depending on the configuration of planetary masses. The resonant argument and the period ratio of the two planets for the fiducial simulation are presented in Fig. \ref{fig:nominalres}.

The three phases of the simulation are clearly distinguishable in Fig. \ref{fig:nominalres}.
Initially, for the first 0.08 Myr, $\psi$ spans the range from $-\pi$ to $\pi$. This circulation happens because the planets are not interacting with each other or the disc, and are initially off-resonance. At the same time, the planet's period ratio stays off resonance and does not evolve since planet-disc interactions are not yet active. Once the planet-planet and planet-disc interactions are activated, the planets migrate into resonance. The resonant argument gets constrained around 0 and the period ratio quickly converges to 1.5, indicative of resonance trapping in the 3:2 resonance. 

As the gap gets more and more depleted as a result of photoevaporation, the interaction between the planets and the gas is affected. The gas is found to act as a damping medium for the eccentricity, as seen in the bottom panel of Fig. \ref{fig:nominalres}, but this damping most importantly limits how deep the resonance can be, or in other words broadens the libration amplitude of the resonant planets. This effect can be seen at later times, where the resonant argument is much tighter constrained in the case of a strong photoevaporative outflow, and the resonance does not show any signs of breaking. A similar behaviour is observed in the period ratio that broadens in the absence of strong photoevaporation. These results indicate that there is a threshold in photoevaporation strength above which the resonance stabilises as a result of the gas depletion in the gap. In the other cases, the resonant argument evolves to larger and larger values with time, until the resonance is inevitably broken. 
In the scenarios where the resonance breaks, the simulation is stopped following a scattering event. For strong photoevaporation, the simulation is stopped after the inner disc is depleted.
Possible resonance breaking mechanisms are discussed in section \ref{paramsec}, where this fiducial simulation is compared with different setups.

There is no significant difference in the evolution of the resonance between cases of weak photoevaporation ($\dot{M}_{\text{PE}} = 10^{-9} \: \mathrm{M_\odot \,yr^{-1}}$ and $\dot{M}_{\text{PE}} = 10^{-10} \: \mathrm{M_\odot \,yr^{-1}}$) and those without photoevaporation at all. In other words, the resonance breaks in the exact same way. This similarity can be explained by the relative strength of two competing effects: gap depletion by photoevaporation and gap refilling by viscous diffusion, a competition we explore further in section \ref{alpha}.

 The eccentricity evolution of the inner planet (which is higher than the outer planet) is shown in Fig. \ref{fig:nominalres}, which provides further evidence for the impact photoevaporation plays on clearing gap material. Since the gap is depleted in the case of $\dot{M}_{\text{PE}} = 10^{-8} \: \mathrm{M_\odot \,yr^{-1}}$, there is limited gas to damp the planet's eccentricity. In this specific simulation, the inner disc also dissipates quickly, so the eccentricity stabilises. The fact that the eccentricity decays at roughly the same rate in the weaker photoevaporation cases after $\sim $0.2 Myr is consistent with the idea that a photoevaporation rate above a certain threshold is required to impact the stability of the resonances. This concept of a threshold is also evidenced  by the surface density in the gap region for different mass-loss rates shown in Fig. \ref{fig:1Dsurfdens}

\subsection{Migration speed}\label{migration_speed}

\begin{figure}
	\includegraphics[width=\columnwidth]{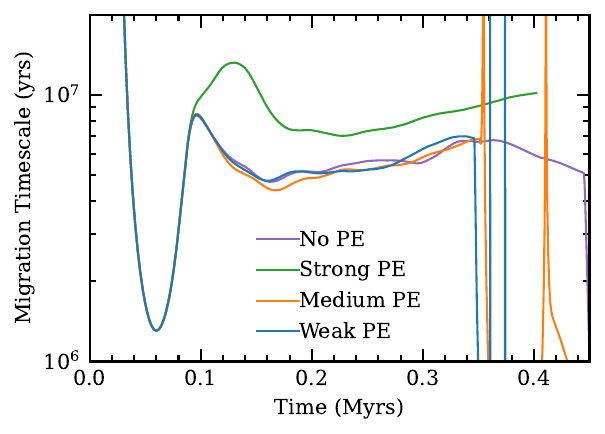}
    \caption{The planetary migration timescale for each wind strength in the nominal simulation. The only difference from the case without photoevaporation is with strong photoevaporation. Before photoevaporation is activated, the migration timescale decreases as the planets migrate quickly into resonance. Once resonance is at its strongest (at 0.1 Myr, see Fig. \ref{fig:nominalres}), the planets migrate in resonance. When strong photoevaporation is operating, the migration speed decreases more initially due to the depletion of the common gap. It then increases again later on due to inner disc depletion. }
    \label{fig:migspd}
\end{figure}

Initial models of type II migration assumed that the gap is fully depleted and that the only element driving migration is viscous diffusion \citep{lin_tidal_1986,syer_satellites_1995}. These models have since been refined, and it is now clear that the material inside and around the gap also impacts the total torque that the disc exerts on the planet \citep{duffell_migration_2014, durmann_migration_2015,kanagawa_radial_2018}. 
Thus, removing material from the gap should also modify the migration timescale. 
To measure this effect, we computed the migration timescale $\tau_{\text{mig}} = {a}/\dot{a}$ from the simulation. Since there are a lot of high-frequency interactions between the planets and with the disc, we smooth out the semi-major axis data by convolving it with a top hat function, hence obtaining a rolling average. This smoothing time was chosen to remove these high frequency oscillations while not affecting the long timescale behaviour. We found a top hat with a time span of 0.0318 Myr gave robust results.  Additionally, since we require $\dot{a}$, we take a window of the same width from the smoothed-out data and find the best fit gradient, by fitting a straight line with a least squares approach. We then compute $\tau_{\text{mig}}$ from this modified dataset, and the results are shown in Fig. \ref{fig:migspd}. The migration timescale is impacted only in the case of strong photoevaporation; weak photoevaporation does not affect the planetary evolution but migration slows down once photoevaporation depletes the common planetary gap. 

\subsection{Parameter exploration} \label{paramsec}

In this section, we investigate how the resonance type, disc mass, viscosity and planet masses impact the disc's and planets' evolution. 

\subsubsection{Resonance type and disc mass}

The 3:2 resonance was chosen for the fiducial model, as it is a common resonance\footnote{It is additionally found to be the most prevalent near-resonant detection for the Kepler sample \citep{fabrycky_architecture_2014}} that is generally dynamically stable and represents a balance between the 2:1 resonance where the planets are furthest away from each other and the other first order resonances that are easier to disrupt.
We now use our simulation set to investigate whether our previous results are also observable in the case of 2:1 resonances, as well as exploring the total disc mass, which changes the relative importance of photoevaporation in the gap. 

\begin{figure*}
	\includegraphics[width=\textwidth]{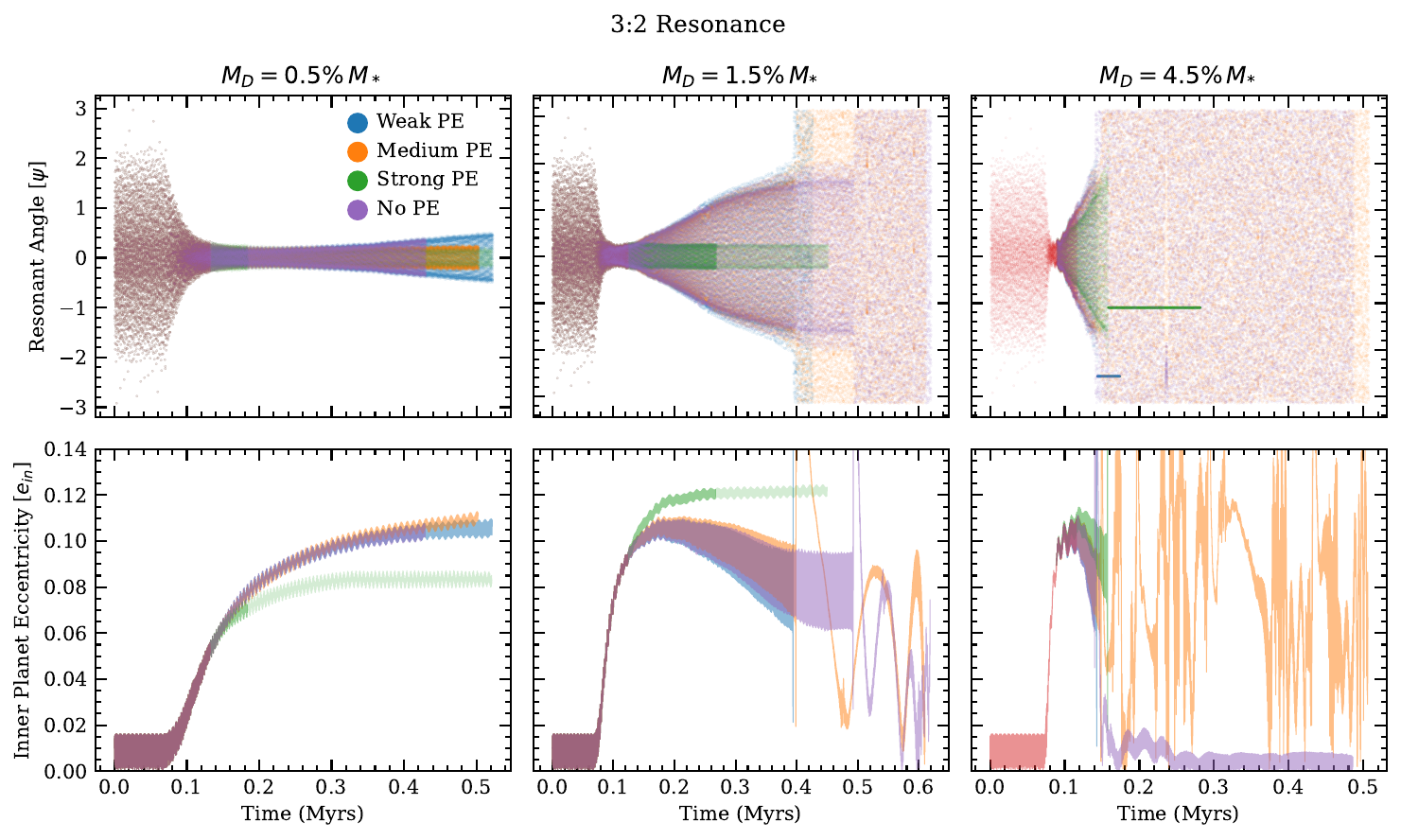}
    \caption{The dynamical evolution for different disc masses for planets in a 3:2 resonance. The middle row represents our fiducial case. Top row: The evolution of the resonant angle as a function of time. The lower the disc mass, the slower the libration amplitude increases. If the resonance breaks too fast, photoevaporation will have no effect. Bottom row: The evolution of the inner planet's eccentricity. Photoevaporation weakens planet-disc interactions, so the eccentricity can increase due to planet-planet interactions. If the disc is less massive, photoevaporation might deplete the gap before the planets fully migrate into resonance, hence the eccentricity might end up lower in this case. The lighter colours indicate that the disc has been depleted beyond the outermost Lindblad resonance of the outer planet.}
    \label{fig:3t2dm}
\end{figure*}

\begin{figure*}
	\includegraphics[width=\textwidth]{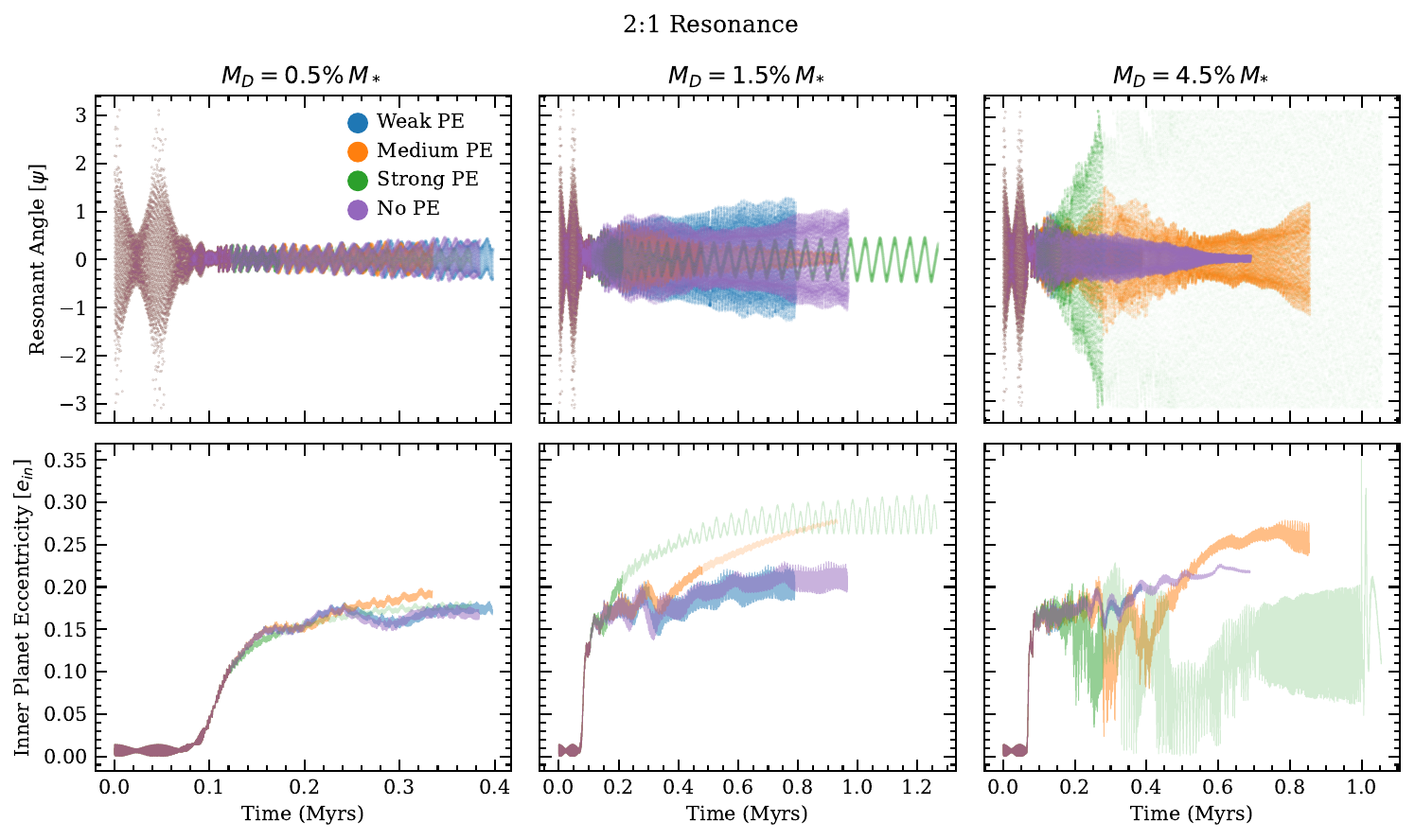}
    \caption{The dynamical evolution for different disc masses for planets in a 2:1 resonance. Top row: the evolution of the resonant angle. In all but once case, the libration amplitude does not seem to significantly increase, leading us to conclude that this resonance is stable on the timescales studied here. For high photoevaporation and high disc masses, the resonance ends up becoming unstable.  Bottom row: the evolution of the inner planet's eccentricity. Similarly to the 3:2 case (Fig. \ref{fig:3t2dm}), low disc masses will not produce higher eccentricities due to photoevaporation but for the middle panel, we see that high photoevaporation can significantly increase eccentricity and even medium photoevaporation exhibits a similar behaviour at later times. This increase in eccentricity is also reflected in the decrease of libration amplitude. The lighter colours indicate that the disc has been depleted beyond the outermost Lindblad resonance of the outer planet.}
    \label{fig:2t1dm}
\end{figure*}

Figs. \ref{fig:3t2dm} and \ref{fig:2t1dm} compare the evolution of the resonant argument and the planetary eccentricity for the 3:2 and 2:1 resonances as a function of disc mass. We see that in the absence of photoevaporation, the resonance lifetime decreases as the disc mass increases. This correlation is especially evident in the case of 3:2 resonances (Fig. \ref{fig:3t2dm}). This result arises because of two different effects; the first one is through planet-disc interactions, where the libration amplitude steadily increases with time, and the resonant angle eventually starts to circulate. However, since the giant planets evolve in a gap, these planet-disc interactions are weaker, and hence the resonance breaks slower than for planets evolving in the type-I migration regime. They hence spend more time with higher eccentricities and large resonant angles, thus increasing the likelihood of planet-planet scattering.

Planet-planet scattering is the second effect leading to resonant breaking. For example, in the fiducial case without photoevaporation, the resonance angle librates at an increasingly large amplitude from the moment the planets get into resonance (at around 0.1 Myr), steadily weakening the resonance until a scattering event at 0.4 Myr completely breaks it. Scattering events are more likely in the 3:2 resonance configuration simply because the orbits are closer together, and this is likely why we observe no scattering events for the 2:1 resonance simulations. The resonance breaking timescale is found to be much longer than the simulation time, so no formal resonance breaking is observed either for 2:1 resonances. This absence of resonant breaking is similarly observed in the low disc mass simulation with planets in 3:2 resonance. In this case, we find the evolution of the maximum libration amplitude by recording the maximum of the resonant argument (presented in the upper left panel of Fig. \ref{fig:3t2dm}) over a rolling window of 0.008 Myr. In the increasing portion of the recorded maximum, we fit a line and find an amplitude growth gradient of roughly $7 \times 10^{-1} \,\text{rad}\, \text{Myr}^{-1} $, which is very slow and implies that at this rate we would need to run the simulation for roughly 5 Myr (i.e 10 times longer), which is not feasible. This gradient is an order of magnitude smaller than the one measured in the fiducial case where the disc mass is $1.5\% \,\text{M}_*$. Hence, the disc mass is an important parameter in breaking giant planet resonances.

In the presence of photoevaporation, we see in the case of a larger disc mass that resonance breaking is marginally affected, because the gap depletion occurs on a timescale longer than that of resonance breaking. There is more material in the gap to deplete, on a shorter timescale due to the fast resonance breaking, so even strong photoevaporation does strongly not impact the resonance breaking timescale, and the planets scatter regardless. 

In contrast, for a lower disc mass, the resonance breaking timescale is longer, and there is less material for photoevaporation to deplete in the gap. Hence, weaker photoevaporation has a stronger impact at low disc masses. However, this impact is mitigated by the fact that the resonance is less likely to break, so even strong photoevaporation makes little difference in this region of parameter space. What could happen is that strong photoevaporation can deplete the inner disc before the eccentricity is fully excited due to convergent migration, leaving us with systems similar to the one in 3:2 resonance and a low disc mass, i.e with a low libration amplitude, but also low eccentricities.

When considering the case of 2:1 resonances in Fig. \ref{fig:2t1dm}, the trends are less clear. The simulations without photoevaporation reveal that this resonance is generally more stable, even at high disc masses, because the resonant argument is typically found to librate. This 2:1 configuration is more stable because the planets are farther apart from each other than for a 3:2 resonance.
Either the 2:1 resonance is stable, or we would need to run the simulation much longer to observe any breaking.

For the lower and intermediate disc masses simulations in  2:1 resonances, we observe the same effect as in the 3:2 resonance case: high photoevaporation rates deplete the common gap making the resonance deeper. 
In the case of the lowest disc mass, we again notice that the gap and inner disc are depleted fast enough to prevent photoevaporation from enhancing the resulting eccentricities. For an intermediate disc mass, we do get a significant increase in the planet's eccentricity owing to strong photoevaporation. We also note this increase at later times with the intermediate photoevaporation strength, once it has had sufficient time to deplete the gap. 

\begin{figure}
	\includegraphics[width=\columnwidth]{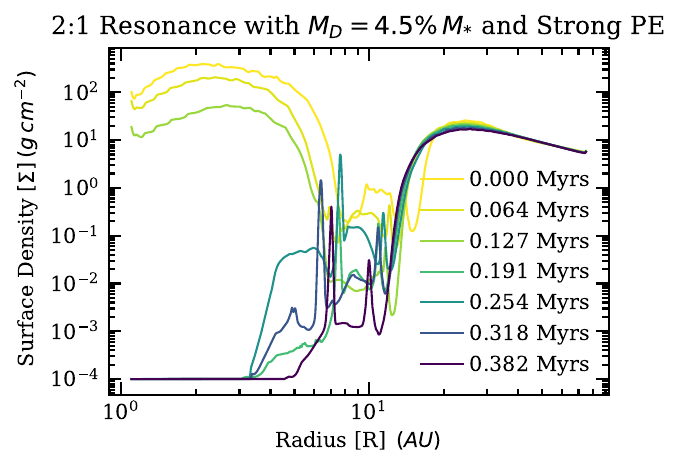}
    \caption{Evolution of surface density for planets in 2:1 resonance embedded in a massive disc with strong photoevaporation. The basic evolution is similar to what was previously discussed, except that the common gap is never fully depleted, due to material flowing from the massive outer disc. This flow means that planet-disc interactions are not completely neutralised, and the resulting surface density distribution, (depleted inner disc, shallow but still existing gap, and unaffected outer disc) leads to resonance breaking.}
    \label{fig:1Dsurfdensbreak}
\end{figure}

At high disc masses, we observe the inverse effect; the stronger the photoevaporation, the more the resonance becomes unstable. This surprising effect can be understood as a consequence of the imbalance in surface density between the inner and outer disc. This imbalance is demonstrated in Fig. \ref{fig:1Dsurfdensbreak}, where the higher surface density in the outer disc allows for more material flowing into the inner disc, which either gets trapped in the gap or can build-up temporarily outside the inner edge. This flow further means that the planet-disc interactions are not uniformly removed across the gap, leading to imbalances that contribute to breaking the resonance gradually over time, without any scattering events. This behaviour is similar to that found in \citet{marzari_shifting_2018}, where the author aims to evaluate the shift in semi-major axis of the resonance due to a disc, and also investigates the stability of these resonances as the disc dissipates, by numerically integrating a three body problem with an additional disc potential. The author finds that resonance breaking is more common as more massive discs dissipate. Here we are able to recover their results, but we also show that the problem is much more complex, since in most cases dissipation does not break resonances. This discrepancy is due to their dissipation model which is simply an exponential decrease of the surface density.

\begin{figure}
	\includegraphics[width=\columnwidth]{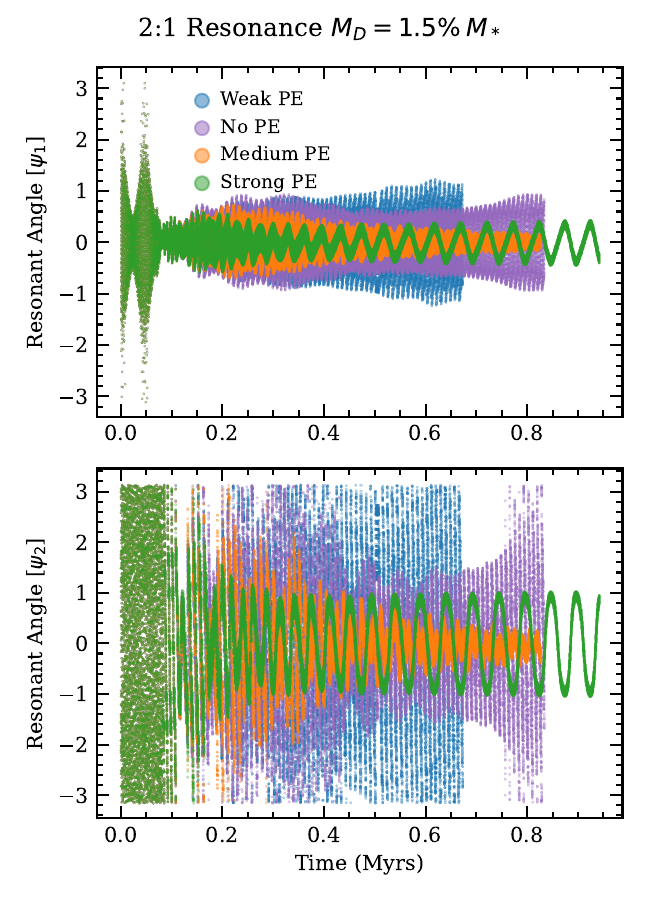}
    \caption{Comparing the two resonant angles' evolution in the case where the planets are in 2:1 resonance with $M_{\text{disc}} = 1.5\% \: M_{*}$, in the presence of different photoevaporation strengths. The first resonant angle librates in all cases and is marginally affected by photoevaporation. However the second resonant angle circulates without photoevaporation. Medium and strong photoevaporation rates lead to the progressive libration of $\psi_2$. This switch happens quicker for higher photoevaporation rates.}
    \label{fig:mixedlib}
\end{figure}

Although the resonant angle ($\psi = \psi_1 = (p+q)\lambda_{2} - p \lambda_1 - q\varpi_1$) presented in Fig. \ref{fig:2t1dm} for the 2:1 resonance clearly librates, to paint a full picture of resonant stability we can look at the second resonant angle ($\psi_2= (p+q)\lambda_{2} - p \lambda_1 - q\varpi_2$), with the difference here being that it is calculated with respect to the (inner/outer) planet's pericentre. As shown in Fig. \ref{fig:mixedlib}, without photoevaporation $\psi_2$ oscillates and librates for a short period of time, before finally circulating again. This discrepancy - the libration of $\psi_1$ and simultaneous circulation of $\psi_2$ - means that the 2:1 resonances as observed here are not as robust as one might originally think. This discrepancy can be further interpreted dynamically as the disc migrating the planets into resonance, but once the resonance is strong enough, it dissipates the eccentricity and breaks the resonance ever so slightly. By including photoevaporation, we notice that $\psi_2$ ends up librating without interruption, after 0.2 Myr for strong photoevaporation and after roughly 0.4 Myr for the medium strength photoevaporation case. In fact, even in the cases without photoevaporation, we observe a tentative return to librations after a certain amount of time, which coincides with the viscous timescale $\tau_{\nu} = {R^2}/{\nu}$ at the inner boundary (1 AU, $\sim4 \times 10^5$ years). However, this effect is not found for weak photoevaporation, so this momentary libration might lead to further circulation and may simply be a result of the stochastic nature of these interactions. Hence we find that dissipation, whether it be from viscous effects or from photoevaporation, removes disc material and leads to resonances becoming more stable through the libration of the two resonant angles instead of only one when the disc is still present.

\subsubsection{Viscosity and Planet mass}  \label{alpha}

\begin{figure}
	\includegraphics[width=\columnwidth]{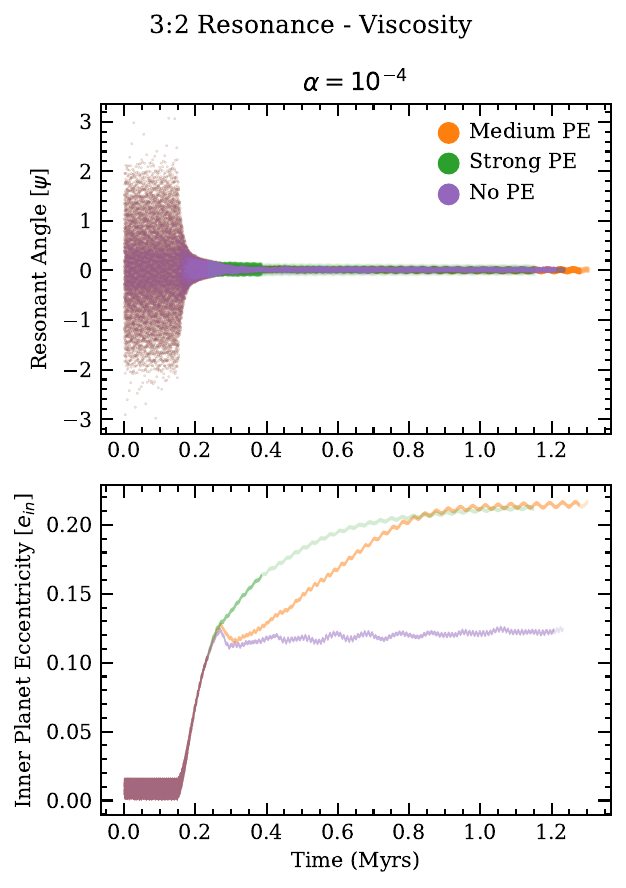}
    \caption{The evolution of the resonant angle (top) and inner planet's eccentricity (bottom) for a disc with a lower viscosity ($\alpha=10^{-4})$, but otherwise identical parameters to the nominal case. The lower viscosity leads to less viscous flows into the gap, hence the resonance is more stable. Photoevaporation has a strong effect as demonstrated by the intermediate photoevaporation raising the eccentricity in comparison to the nominal case with $\alpha=10^{-3}$ (Fig.~\ref{fig:nominalres}). The lighter colours indicate that the disc has been depleted beyond the outermost Lindblad resonance of the outer planet.}
    \label{fig:3t2a4}
\end{figure}

 In Fig. \ref{fig:3t2a4} we took the fiducial setup and ran it with a weaker viscosity of $\alpha = 10^{-4}$, as this might be more realistic of the turbulence level in a wind-driven disc \citep[e.g.][]{rosotti_empirical_2023, villenave_turbulence_2025}. 

A weaker viscosity naturally results in deeper gaps \citep{duffell_gap_2013}, and hence weaker planet-disc interactions in the gap region, which naturally results in more stable resonances. Our simulations demonstrate this, where we see a smaller libration amplitude of $\psi$ in the top panel of Fig. \ref{fig:3t2a4}. We first consider the case without photoevaporation. Comparing this case to the fiducial setup ($\alpha = 10^{-3}$) which immediately shows signs of resonance breaking, this setup ($\alpha = 10^{-4}$) shows the opposite, i.e. resonance gets stronger because the resonant argument librates with a decreasing amplitude. So, a weaker viscosity may favour the development of resonances in giant planet systems. 
Since here the planet-disc interactions favour resonance, minimising them through photoevaporation leads to the resonant argument being unable to reach smaller amplitudes. These same interactions are still responsible for damping out the eccentricity, so photoevaporation still leads to higher eccentricities (around a factor of 2 in the most extreme case). The weaker viscosity also allows the intermediate photoevaporation strengths to have a more significant effect, especially on the inner planet's eccentricity.

\begin{figure*}
	\includegraphics[width=\textwidth]{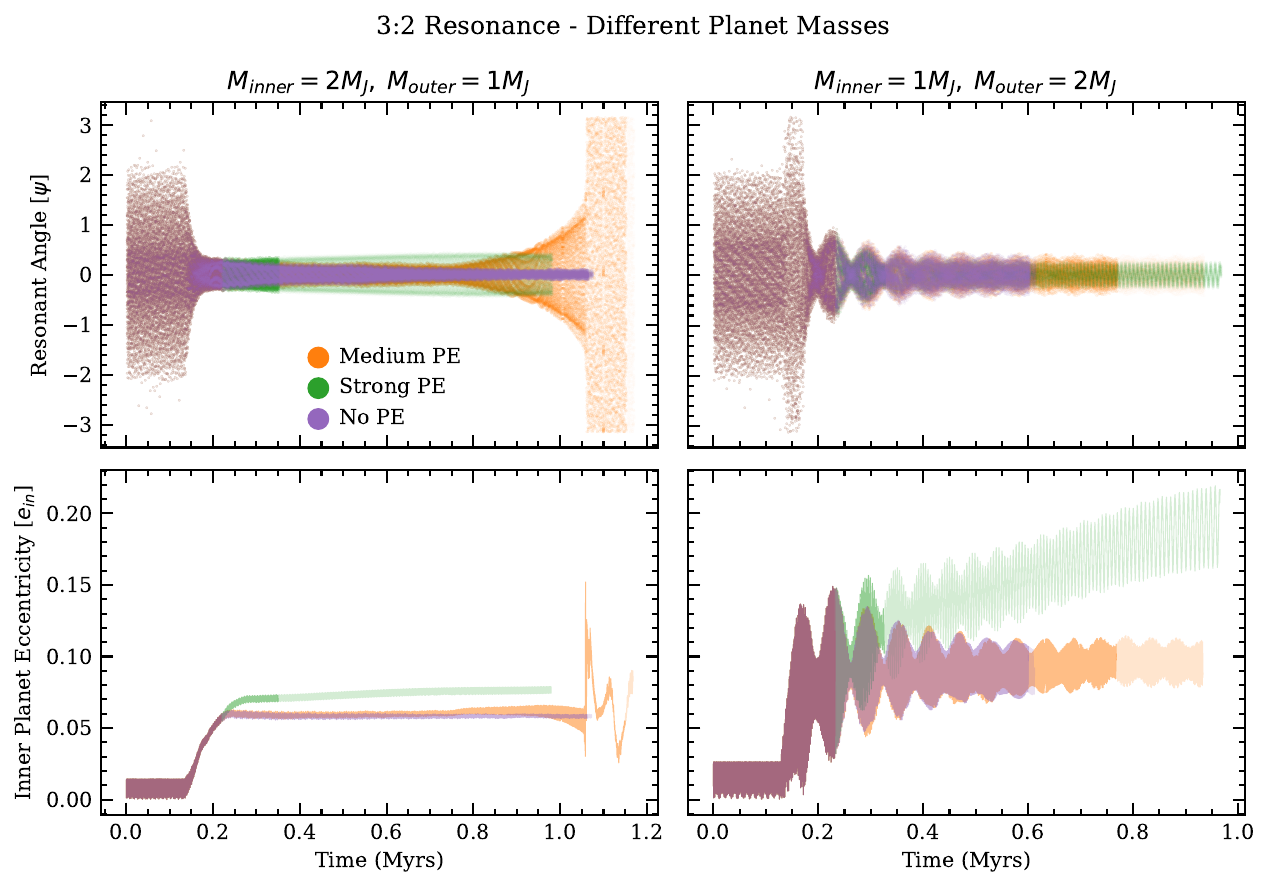}
    \caption{The evolution of the resonant angle (top) and inner planet's eccentricity (bottom). The simulation parameters are identical to the nominal case apart for the planet masses. First column: placing a more massive inner planet leads to stable resonances, with an increase in eccentricity. The resonance is more robust without photoevaporation, since photoevaporation prevents the planets from fully migrating into resonance, despite larger eccentricities. Resonance breaking is seen for the intermediate photoevaporation around 1 Myr, possibly reminiscent of the behaviour of the 2:1 strong photoevaporation massive disc case. Second column: making the outer planet more massive also leads to stable resonances, with no impact from photoevaporation, apart from an increase in the inner planet's eccentricity as seen in the nominal case. The lighter colours indicate that the disc has been depleted beyond the outermost Lindblad resonance of the outer planet.}
    \label{fig:3t2mpl}
\end{figure*}

We also separately changed the fiducial model to run it with uneven planet masses, leading to a simulation with a $2 M_\text{J}$ inner planet and another simulation with a $2 M_\text{J}$ outer planet, as shown in Fig. \ref{fig:3t2mpl}, while the other planet's mass was held at 1 M$_{\rm J}$. 

In varying the planet masses, we confirm that the fiducial setup, with equal masses, is the most unstable, since these two setups have stable resonances on Myr timescales. In Fig. \ref{fig:3t2mpl} we show the general absence of any resonance breaking without photoevaporation. Similarly to the previous case, planet-disc interactions strengthen resonance; therefore, photoevaporation does not produce deeper resonances, though high viscosity means that the eccentricity is overall lower. There is, however, a clear sign of resonance breaking for the medium photoevaporation in the case of a more massive inner planet, after about 1 Myr. This breaking is a similar effect to that discussed for the heavy 2:1 resonance, where partial depletion of the gap can lead to torque imbalances.

\section{Discussion}

The goal of this study was to use hydrodynamical simulations to investigate the long-term impact of disc photoevaporation on the evolution of mean motion resonances between giant planets. In this section, we discuss the assumptions that we made in carrying out this work and how they may impact our results, as well as discussing the wider context and implications of our results. 

\subsection{Disc and Planet Parameterisation} \label{parameterisation_discussion}

When setting up the initial disc structure and properties, certain assumptions were made. We used a constant $\alpha$ viscosity. The physics behind the determination of $\alpha$ is complex, as it may include magnetic fields and the impact of dust on the ionization of the gas. \citet{delage_impact_2023} explores this problem deeper and finds that effective $\alpha$ might have a steep spatial variation around 10 AU, going from $10^{-3}$ to $10^{-4}$. Given that the disruption to the resonances varies significantly between these two viscosity values in our study, this variation might significantly impact resonance breaking in real discs. Another factor to consider is the impact of the planets and gaps themselves on the dust profile, and therefore this might lead to time dependent viscosities.

Another assumption of our model is the adoption of a locally isothermal disc profile. This model neglects variations to the thermodynamic profile within the gap, and hence can only approximate the migration behaviour. \citet{ziampras_modelling_2023} investigates these issues and finds that in the case of a low viscosity disc, implementing radiative transfer in the simulations the main gap structure does not change significantly. This is because in the outer disc, the thermal timescales are much smaller than the dynamical timescale. 
Additionally, the relatively higher viscosity used in our study is responsible for smoothing out any changes due to shocks, and the extreme depletion in the gap will have a more significant effect in changing any remaining Lindblad torques that the marginal changes from the thermal conditions, so the isothermal condition is still appropriate here and we argue that the inclusion of more detailed radiative transfer will not have a major impact on the outcome of our simulations.
A lower value of $\alpha$ might also be more appropriate, as recent observations \citep{trapman_observed_2020} have found that they might be better suited to explain the structures found in disc observations by ALMA. 

Lower values of $\alpha$ also impact the stability of resonances, as shown by \citet{griveaud_migration_2023}. The authors find that in low viscosity discs ($\alpha = 10^{-4}$), the outer planet can never reach the migration speed necessary to cross the 2:1 resonance. We find in this study that the depletion of the inner disc can lead to an asymmetry in the torques, as shown in Fig. \ref{fig:1Dsurfdensbreak}, which facilitates resonance breaking. Although our results concern simulations with $\alpha = 10^{-3}$, we expect that this modified density profile might also lead to 2:1 resonances breaking in lower viscosity discs.

The planets, as modelled here, are non-accreting, which also impacts their migration rate and hence the overall planet-disc interactions. If the planet is allowed to accrete, then less material crosses the gap, and therefore the migration rates are lower, as found in \citet{durmann_accretion_2017} and \citet{li_concurrent_2024}. We could therefore reasonably expect the resonances to be more robust if planet accretion were allowed. However, here we model a system in the late stages of its lifetime, and we can assume that the planets accretion rate is small. Additionally, strong photoevaporation might remove most of the gas before it reaches the planet, so the non-accreting regime seems more applicable in our case.

The disc's scale height also plays an important role in defining the depth of the gap and its sharpness \citep[e.g][]{kanagawa_formation_2015}. Many simulations tend to run unrealistically large values of $H/R$, for the sake of computational efficiency. Our values are consistent with a passive disc, allowing us to calculate the scale height from the temperature profile. These values are consistent with the observational results, which strengthens the assumption \citep{andrews_protoplanetary_2010}. Lower values of scale height, as prescribed in this study, lead to deeper, sharper gaps and hence slower migration and slower resonance breaking. Furthermore, these deeper gaps are the perfect environment for disc photoevaporation to have a greater influence on the disc profile, as it can more quickly deplete the common gap. 
To demonstrate this reasoning, we ran models with a larger scale height and compare them to our fiducial model in Appendix \ref{scaleheight3}, confirming that photoevaporation is more important for the more realistic discs simulated in this work. Additionally, we find that the dynamical state of the planets is greatly influenced by the choice of scale height.

Finally, to allow for long-term evolution, we needed to resort to using a 2D disc model. 2D models affect not only the overall disc dynamics, but also require a smoothing potential (Eq. \ref{smoothing}), which can affect the magnitude of the torque on the planet. \citep{masset_co-orbital_2002}. \citet{brown_horseshoes_2024} have shown that the smoothing potential used here and in most hydrodynamical simulations can overestimate the one-sided Lindblad torques by a factor of 3, compared to the full 3D treatment of the problem, as also emphasised later this motivates further study of our problem with 3D simulations. To check that our results are independent of the smoothing length used, we ran simulations with different 2D smoothing lengths, presented in Appendix \ref{smoothing_BO}. We find little to no difference in the qualitative evolution of the resonance locking. We also include the 2D prescription from \citep{brown_horseshoes_2024} to compare, and also find no qualitative difference

\subsection{Choice of photoevaporation model} \label{PEmodelchoice}

\begin{figure}
    \includegraphics[width=\columnwidth]{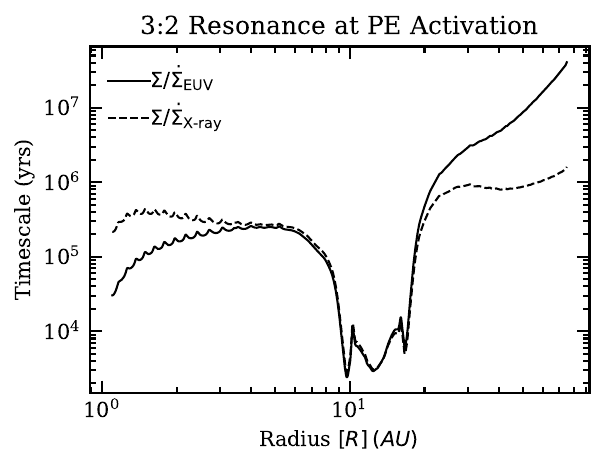}    
	\includegraphics[width=\columnwidth]{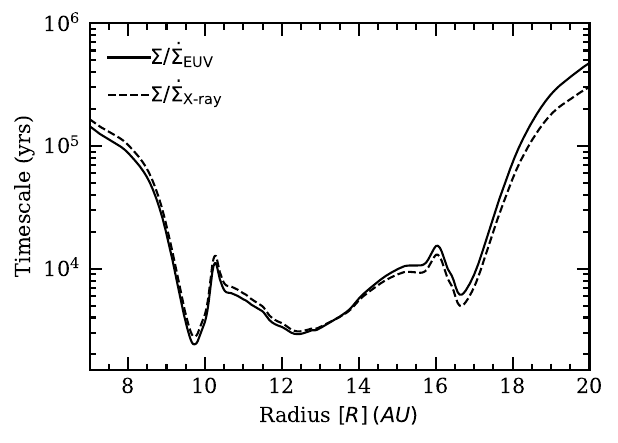}    
    \caption{Comparing the local photoevaporative depletion timescales ($\Sigma/\dot{\Sigma}_{\text{PE}}$) in the common gap for EUV and X-ray \citep{Owen2012} photoevaporation, where the surface density profile is extracted from the fiducial simulation before photoevaporation is activated. The surface density mass loss rates are matched in the middle of the gap, and we see that the depletion timescale varies negligibly between the two profiles in the gap.
    }
    \label{fig:timescales_xray}
\end{figure}

This study aimed to evaluate the impact of photoevaporation on only a small area of the disc (i.e., in the common planetary gap), so the exact global shape of the model used here is less relevant to our problem than other work on the impact of photoevaporation on disc evolution. Thus, we expect resonances to be more robust with strong photoevaporation regardless of the irradiation type, based on the general principle that disc mass loss in the common gap reduces the dynamical disruptions to the resonance. Although running an EUV model with $\dot{M} = 10^{-8} \mathrm{M_{\odot} \, yr^{-1}}$ is an overestimate for this specific mass-loss model, these mass loss rates can be achieved in other regimes (FUV and X-ray). Fig. \ref{fig:timescales_xray} compares the depletion timescales for EUV and X-ray photoevaporation in the gap region, for an equivalent surface density mass loss rate in the middle of the gap. We find that the radial variations in each profile do not significantly affect the photoevaporation timescale across the gap. The main difference stems from the X-ray profile's longer depletion timescale in the inner disc and shorter depletion timescale in the outer disc, leading to a possible faster clearing of the outer disc compared to the simulations presented here.

Therefore, since we are only focused on the local aspect of photoevaporation in the common gap, it is appropriate to use the simple EUV profiles which we scale up and down to values appropriate for all photoevaporation models. \citet{jennings_comparative_2018} used a similar motivation to justify the physical relevance of high EUV photoevaporation strengths in their comparative approach to the impact of photoevaporation on the evolution of single giant planets. However, \citet{jennings_comparative_2018} did find that the photoevaporation model can affect the long-term statistical distribution of planets as they migrate closer to $R_\text{g}$. In our work, we have not attempted to run a statistical study and are only concerned with our parameter study, which is focused on the physics of resonance evolution and breaking.

Additionally, the EUV model we use from \citet{alexander_dust_2007} is only the optically thick, ``indirect'' model where the inner disc is not yet depleted. In some of our simulations, we extend the run beyond inner disc depletion, where the ``direct'' model would be more appropriate. In this regime the photoevaporative outflow originates from the photons directly hitting the disc edge (instead of recombination photons from the upper disc's atmosphere). Hence, the outer disc should be removed much faster than it is in our standard simulations after inner disc dispersal.

To find the parts of the planetary orbital evolution that are affected by the limitations imposed by the direct irradiation regime, we highlight the point at which the inner disc is depleted with a lighter shade of colour in Figs. \ref{fig:3t2dm}, \ref{fig:2t1dm}, \ref{fig:3t2a4} and \ref{fig:3t2mpl}. We find this depletion time by first finding the moment at which the inner 1.5 AU of the disc is depleted to the floor value. We then find the inner radius of the cavity by searching for the outermost point in the disc where the surface density is less than the threshold value, set to $10^{-1}\: \mathrm{g~cm^{-2}}$. We use this inner radius to calculate the direct irradiation mass loss profile from \citep{alexander_dust_2007}, and estimate the time it would take to completely deplete the disc up to the outermost Lindblad resonance with the outer planet. To calculate this timescale, we take the ratio of the disc mass contained within this outer limit to the integrated direct photoevaporation rate up to the same outer limit. We hence expect there to be no significant dynamical contributions of the disc to the planets' evolution beyond this point in time.

We find that all strong photoevaporation simulations get depleted quickly, except for the high disc mass setup with planets in 3:2 resonance, where the planets scatter before photoevaporation is able to create a cavity in the disc. We also invoked massive inflows from the outer disc to explain why the resonance is unstable in the model with 2:1 resonances, high disc mass, and strong photoevaporation. If the outer disc gets depleted, these mass flows would perturb the system less, perhaps limiting the resonance breaking, since it is still happening on a relatively long (0.1 Myr) timescale. Fig. \ref{fig:2t1dm} shows, however, that the direct regime becomes relevant after the resonance is already broken, so our findings concerning resonance breaking due to photoevaporation are unaffected by this limitation.

In some cases the depletion occurs while the eccentricity is rising. We therefore expect the eccentricity to plateau at lower values than what was observed in these simulations. In the specific case of Fig. \ref{fig:3t2a4}, the eccentricity was found to plateau at the same critical value for both strong and medium photoevaporation; however, we might expect the value of this plateau to be lower for strong photoevaporation, given that the depletion occurs before the plateau.
Given that the direct irradiation regime is only relevant once the inner disc and the planetary gap are depleted, and the short timescales over which this depletion occurs, we do not expect it to contribute significantly to gap clearing. Given also that the direct regime only activates once planet-disc interactions are already weak, we do not expect this regime to significantly alter the torque balance and hence the process of resonance formation and breaking. 
However, we must caveat this with the statement that the large difference between indirect and direct photoevaporation is only applicable to EUV driven photoevaporation and is not so large in X-ray driven photoevaporation \citep[e.g.][]{Owen2012,picogna_dispersal_2019}. 

In addition to photoevaporation, magnetohydrodynamic (MHD) effects can generate a wind in the disc. Although its effect on resonance breaking has not yet been researched, 3D simulations with a single planet have found that including MHD winds also results in deeper gaps \citep{aoyama_three-dimensional_2023}, possibly reducing planet-disc interactions that lead to resonance breaking. When it comes to migration, 2D models with wind prescriptions \citep{kimmig_effect_2020} found that MHD winds can lead to strong outward migration. 3D models later found that migration can be directed inward \citep{aoyama_three-dimensional_2023} or outward \citep{wafflard-fernandez_gone_2025}, so uncertainty still remains on how the MHD wind would affect the differential torques causing the planets to migrate in or out of resonance.

\subsection{Impact of photoevaporation on the planetary gap}

The photoevaporation model used here is azimuthally symmetric and does not respond to changes in the disc structure, and as such is constant in time. \citet{Owen2012} argued that photoevaporation rates should be relatively invariant to the structure of the disc, which should adjust their flows to feed the photoevaporative outflow rather than vice versa.
However, \citet{weber_interplay_2024} included the \citet{picogna_dispersal_2019} parameterisation in a 3D simulation with a single giant planet. They found that a pressure gradient inversion in the photoevaporative outflow leads to the single planet's gap being filled with material leaving the disc just outside the gap. They concluded that the picture of deeper planetary gaps arising from photoevaporation should be revised, and hence we should be cautious about our claims about photoevaporation producing stable resonances.

However, we expect that the setup investigated here may not be subject to the refilling effect observed by \citet{weber_interplay_2024}. The major difference here is that we are looking at multiple planets in a common gap, so the photoevaporative inflow will have a lesser impact on the $\sim R$ ($\sim$ 10AU in our case) wide common gap, compared to the $\sim H$ ($\sim$ 2AU) wide gap in their study. Additionally, their simulation used a larger $H/R$, resulting in an increased forcing of photoevaporative outflow material towards the gap. Their planet was also closer in, where it could build a deeper gap ($\Sigma_{\text{gap}}/\Sigma_0 = 10^{-4}$, versus our $\Sigma_{\text{gap}}/\Sigma_0 = 10^{-2}$).  Hence, we suspect that while photoevaporation may refill single planetary gaps in close orbits as seen in \citet{weber_interplay_2024}, it would be unable to impact the large, $\sim R$, common gaps relevant for multiple resonant giant planets.

\subsection{Eccentricity damping}

In this study, we find that the eccentricity damping of the giant planets correlates strongly with resonance breaking. Here, we discuss possible physical mechanisms for this damping. \citet{teyssandier_growth_2016} develop a linear model of eccentricity damping of giant planets in a cavity. They find that viscous and corotation torques contribute to damping the eccentricity, whereas resonant lindblad torques excite eccentricity. Hence, in the cases where the corotation region is depleted, less damping occurs.
Thinking of eccentricity damping as a result of angular momentum exchange with the disc, we notice the similarities between the eccentricity evolution in our simulations (specifically the 3:2 fiducial ones) and those of \citet{ragusa_eccentricity_2018}, where they considered the long-term impact of the disc on a single gap-opening planet. A low mass disc will raise the eccentricity on a timescale of roughly $10^{5}$ orbits to a value of about 0.1 and a heavier disc will first increase the eccentricity on a very short timescale and then damp it away over $10^{5}$ orbits. We also note similar oscillations in eccentricity, due to secular interactions. Since eccentricity rises at least an order of magnitude faster in our simulations, we can infer that planet-planet interactions due to migration are the dominant mechanism for exciting the eccentricity, instead of angular momentum exchange between the planets and the disc.

The origin of eccentricity damping in resonant giant planet-disc interactions is still not fully explored, but since photoevaporation changes the evolution path of eccentricity, we will briefly put these effects into context here. In \citet{lee_dynamics_2002}, the authors model the evolution of the resonant GJ 876 system, and to allow the system to evolve to its current configuration, they introduce an artificial eccentricity damping factor $K$ in their secular model, where $ \dot{e}_\text{i}/e_\text{i}= - K |\dot{a}_\text{i}/a_\text{i}| $ and find that the best models require K=10 or K=100, depending on the migration prescription on each planet. Furthermore, \citet{nagpal_breaking_2024} run a series of N-body simulations of giant planets with different eccentricity damping strengths and compare their results with observations. They find that models with weak damping (K=1-10) are preferred, but they do not further discuss the physical origin of this low damping. We find here that photoevaporative outflows lead to weaker damping and hence higher eccentricities, which is an important effect in characterising how the planets fall into resonance. 

\subsection{Origin of resonance breaking}

Although eccentricity damping from the disc strongly correlates with resonance breaking, it may not be the cause of resonance breaking. \citet{baruteau_disk-planets_2013} run models of partial-gap-opening planets in the inner disc and run models of convergent migration without planet-planet interactions, hence without any significant eccentricity. They still find that once the period ratio is close to that of a mean motion resonance,
even without librations, it slowly diverges away from this value. They find similar results when they include planet-planet interactions. They explain this result by considering the wake from each planet as it crosses the other planet's co-orbital region, hence exchanging angular momentum. This exchange is associated with energy dissipation at constant angular momentum, which leads to divergent migration \citep{lithwick_resonant_2012,batygin_dissipative_2012}. As seen in Fig. \ref{fig:2Dsurfdens}, without photoevaporation the wakes are stronger, and hence resonant repulsion is stronger. In the panel with photoevaporation, there is less angular momentum exchanged, which could be stabilising the resonance. Angular momentum exchange through the other planet's wake may also explain why, in the case of massive discs with significant material in the gap, resonances tend to break more easily. \citet{baruteau_disk-planets_2013} also run simplistic models of disc dispersal right before resonance breaking. Our findings, using more realistic dispersal models and focussing on the outer disc, confirm their results; depleting the gap reduces resonance breaking. However, one difference is that when there is a common gap, they see no effect from disc dispersal. This difference can be explained
because they do not model disc dispersal with photoevaporation, rather they just smoothly deplete the disc globally. We find in our study that dissipation has a lesser effect in the case of a 2:1 resonance than in the case of a 3:2 resonance, where the planets are on average closer to each other and the wakes can interact more. The setup with a common gap used in \citet{baruteau_disk-planets_2013} led the planets' period ratio to approach that of a 5:3 MMR, without librations. This resonance is a second-order one, and trapping leads to much smaller increases in eccentricity in this case. Hence, the planets do not have the opportunity to get closer to one another, so wake interactions are weaker, hence depleting the gap in this case makes no difference, similar to our 2:1 setups.

\subsection{Implications for the search for planets with Gaia}

This work is also relevant to giant planet detections through astrometry, as provided by the upcoming Gaia mission data releases. The masses of the planets and the semi-major axes studied here coincide directly with the detection space of Gaia \citep{perryman_astrometric_2014,ranalli_astrometry_2018}. Having multiple planets within a system could lead to different recovery capabilities, since the astrometric signal of the two planets can overlap and make the combined signal harder to decipher \citep{casertano_double-blind_2008}. However, a feature that has been poorly explored is the impact of resonant configurations on detectability. In \citet{wu_detecting_2016}, the authors investigate what parameter space allows resonances to be better recovered through astrometry, but the problem has not yet been fully studied. Astrometry will provide full orbital elements for many detected planets, allowing the possibility to study their eccentricities and possible resonant configurations, placing useful constraints on our models.

\section{Summary}

We used the hydrodynamical simulation code FARGO3D along with a prescription for disc photoevaporation to investigate the long-term evolution of giant planet resonances in the outer disc regions (at $\sim$ 10 AU). We explored these processes through a parameter space covering different disc masses, viscosities, planet masses and resonances.

The main impact of photoevaporation on resonances comes through its effect on the disc. Photoevaporation is more notable in a local environment where the disc surface density is low; therefore, the effect of photoevaporation is first seen in the common gap from two planets in resonance. By removing material in the gap region faster than it can be resupplied from the gap edges through viscous diffusion, photoevaporation weakens the planet-disc interactions. We find that there is a threshold in the photoevaporation strength above which this effect comes into play; otherwise viscous diffusion refills the gap, where this threshold depends on the disc's parameters.
Our simulations also show longer migration timescales when photoevaporation of the common gap is important as a result of reduced planet-disc interactions. In some cases, the migration timescale can be a factor of two times longer. 

We found that the stability of a 3:2 resonance for equal planet masses depends on the total disc mass. For larger disc masses ($4.5\% \: M_{*}$), the resonance breaks on a time scale of 0.1 Myr, which is too short for even the strongest of our photoevaporation models to have any impact. In contrast, at lower disc masses ($0.5\% \: M_{*}$), the resonance did not show signs of breaking over the 0.5 Myr of simulation time. Here, photoevaporation does indeed reduce the resonant angle libration amplitude but we find it to have little to no impact since the resonances are already stable.

When investigating 2:1 resonances, we found that they are already stable and show no sign of breaking on the timescales studied here. In the case of lower disc masses, photoevaporation also reduces the libration amplitude of the main resonant angle. When considering a more massive disc, we find that strong photoevaporation actually destabilises the resonance once the inner disc is depleted. This destabilisation was tentatively explained by considering that there is still a significant mass flux across the gap because of the massive outer disc. Once the inner disc is depleted, the torque imbalance leads to divergent evolution away from resonance. 

Finally, considering a weaker viscosity ($\alpha = 10^{-4}$) leads to a resonant angle that librates with a very small amplitude in all cases, since the gap is deeper and hence the planets are less perturbed by all the aforementioned effects.  

Overall, the impact of photoevaporation on multiple migrating planets is multifaceted, with no simple straightforward answer. The impact is highly dependent on the parameters of the system, and hence observations are needed to better inform us of which area of parameter space is the closest to reality.
This work is especially relevant in the context of observations of giant planets at separations of around 10 AU. Planet-planet resonant interactions are one possible explanation for some of the high eccentricities observed in the limited sample distant exoplanet systems observed to date. More detailed statistics are expected from Gaia DR4, which will probe these large separations and provide us with a greater characterisation of the architectural properties of cool giant planets.

\section*{Acknowledgements}
We thank the referee, Barbara Ercolano, for  a useful review that improved this manuscript. We also thank Anna Penzlin and Alexandros Ziampras for their scientific insight and feedback. This project has received funding from the European Research Council (ERC) under the European Union’s Horizon 2020 research and innovation programme (Grant agreement No. 853022). EG has received support from an STFC PhD studentship. JEO is supported by a Royal Society University Research Fellowship. This work used the DiRAC Data Intensive service (CSD3) at the University of Cambridge, managed by the University of Cambridge University Information Services on behalf of the STFC DiRAC HPC Facility (www.dirac.ac.uk). The DiRAC component of CSD3 at Cambridge was funded by BEIS, UKRI and STFC capital funding and STFC operations grants. DiRAC is part of the UKRI Digital Research Infrastructure.


\section*{Data Availability}

Data from our numerical models are available upon reasonable request to the corresponding author. The FARGO3D code is publicly available at \url{https://github.com/FARGO3D/fargo3d}.



\bibliographystyle{mnras}
\bibliography{paper1} 




\appendix

\section{The impact of scale height on resonance trapping and breaking} \label{scaleheight3}

\begin{figure}
	\includegraphics[width=\columnwidth]{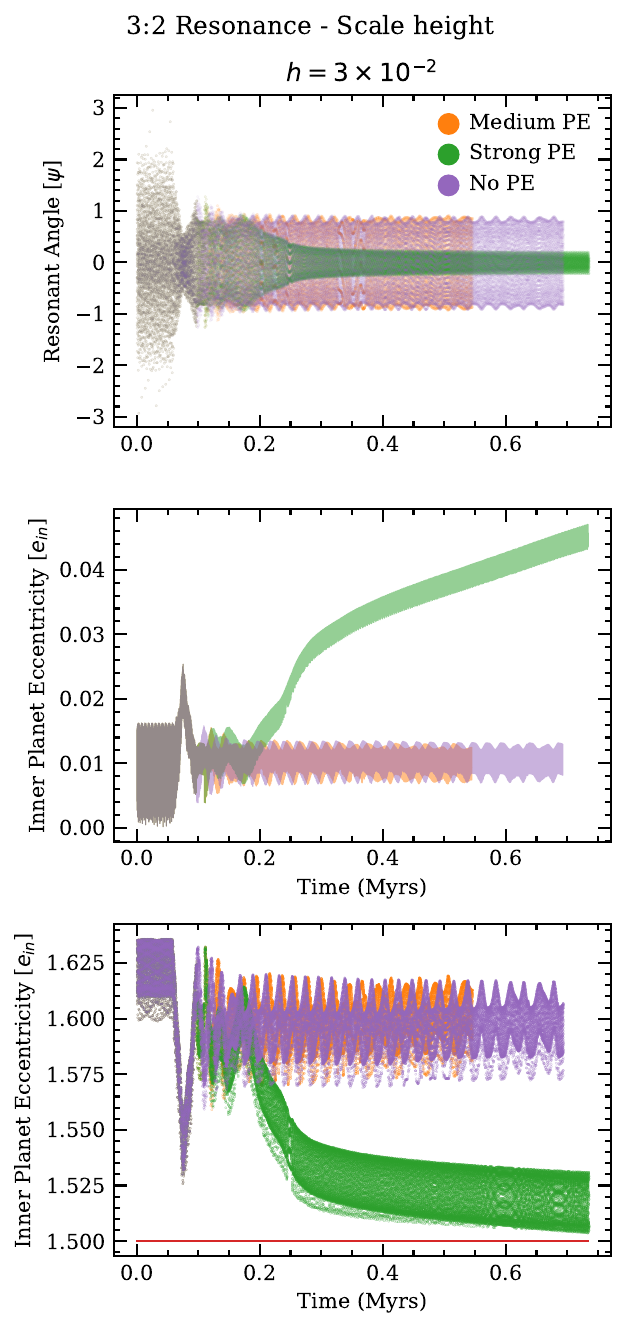}
    \caption{The evolution of the resonant angle (top), orbital period ratio (middle) and inner planet's eccentricity (bottom) for a disc with a higher scale height ($h_0=0.03$) compared to our standard choice, but otherwise identical parameters to the nominal case. The period ratio is further from commensurability in this case, leading to large (stable) librations and low planetary eccentricities. Photoevaporation brings the planets closer to commensurability, resulting in higher eccentricities and smaller libration amplitudes.}
    \label{fig:3t2hr3}
\end{figure}

Many previous studies involving hydrodynamical simulations of discs use unrealistically large values of scale height, for the sake of computational efficiency. We present in Fig. \ref{fig:3t2hr3} the result of the fiducial simulation, with a larger scale height of $h_0 = 0.03$. We find that the resonant angle does increase once the planet-planet interactions are activated, but it stays capped instead of increasing until planet-planet scattering breaks the resonance, as in Fig. \ref{fig:nominalres}. These results show that larger pressures associated with thicker discs result in the planets settling in resonance within a period ratio that is further from commensurability. This shift weakens the planet-planet interactions that lead to the eccentricity increasing, hence the planets achieve lower values of eccentricity in these larger scale height simulations. When considering photoevaporation, we find that the planets are brought closer to commensurability, their eccentricities increase and the libration amplitude gets smaller. We suggest from this result that giant planets found closer to commensurability can be explained by strong photoevaporation, and encourage future work to search for correlations between stellar irradiation and offset from resonance in the future Gaia data.

\section{Planetary potential smoothing} \label{smoothing_BO}

\begin{figure}
	\includegraphics[width=\columnwidth]{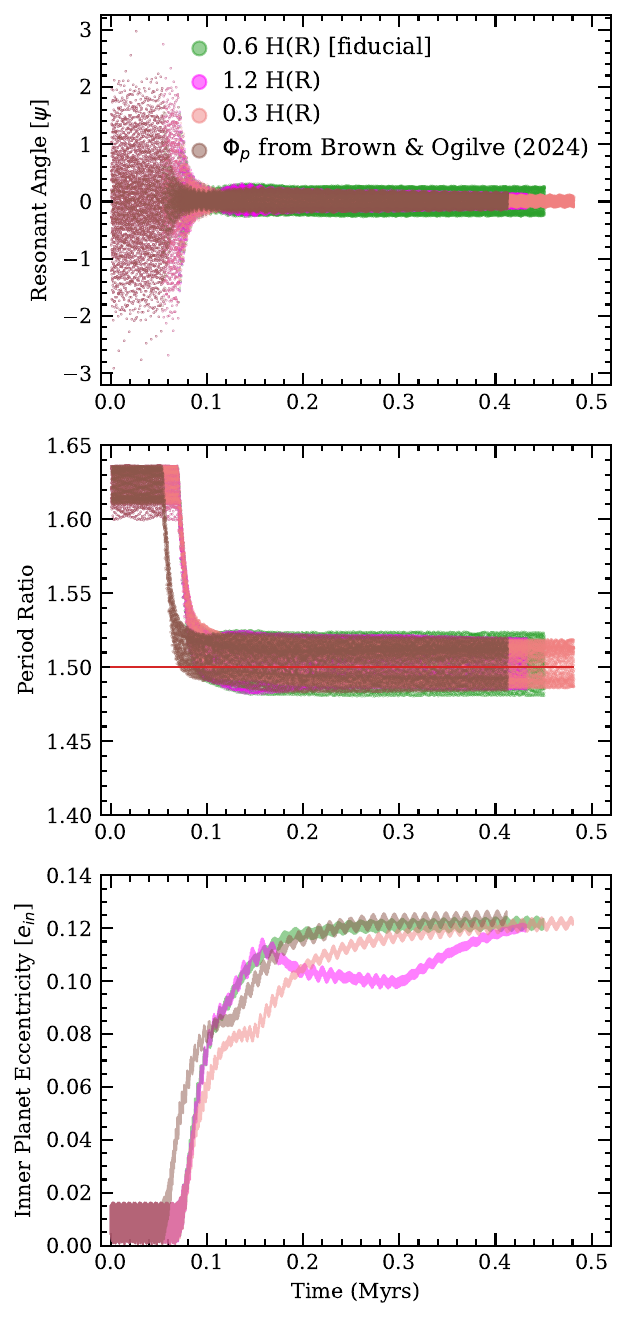}
    \caption{The evolution of the resonant angle (top), orbital period ratio (middle) and inner planet's eccentricity (bottom) for a disc with smoothing lengths $s=0.3 \text{H}$, $s=0.6 \text{H}$ (used in all other runs), $s=1.2 \text{H}$, but otherwise identical parameters to the nominal case, including strong photoevaporation. The smoothing prescription from \citet{brown_horseshoes_2024} is also implemented and included for comparison. All evolution tracks are qualitatively similar.}
    \label{fig:smoothing}
\end{figure}

In section \ref{parameterisation_discussion}, we briefly discuss the limitations that arise from using the smoothing parameter $s$ from Eq. \ref{smoothing}. To make sure that our choice of $s=0.6 H(R)$ does not influence the process of resonance trapping and breaking, we ran the fiducial model again with strong photoevaporation, and use values of $s=0.3 H(R)$ and $s=1.2 H(R)$. The results are shown in Fig. \ref{fig:smoothing}, and we confirm that the smaller smoothing length has qualitatively similar results to the fiducial model. It will result in a lower eccentricity plateau before activating photoevaporation, correlating with the planets' period ratio shift away from commensurability. After photoevaporation is activated, the eccentricity rises and plateaus at the same value as the other smoothing length values, while at the same time the period ratio adjusts itself closer to exact commensurability. The longer smoothing length is also qualitatively similar, stalling the eccentricity at the same value, but the different behaviour (i.e the eccentricity decreasing and increasing back) correlates with the depletion of the common gap  (at 0.15 Myrs) and then the depletion of the inner disc (at 0.3 Myrs).

We also implement the smoothing prescription from \citep{brown_horseshoes_2024}

\begin{equation}
    \Phi_\text{p} = - \frac{Gm_\text{p}}{H} \frac{e^{\frac{1}{4}{s_{\text{b}}}^2}}{\sqrt{2 \pi}} K_\text{0} \left(\frac{1}{4}s_{\text{b}}^2\right), \qquad s_{\text{b}}^2 = \frac{\epsilon^2}{H^2} +  \frac{| \mathbf{r}_{\text{d}} - \mathbf{r}_\text{p}|^2}{H^2},
\end{equation}
where $K_\text{0}$ is the modified Bessel function, $\mathbf{r}_{\text{d}}$ is the position of the gas parcel in the disc, $\mathbf{r}_\text{p}$ is the position of the planet, and $\epsilon$ is defined here to be 0.06 AU, corresponding to the size of a radial cell at 20 AU. We find that this prescription also makes no difference to the global evolution of resonances. Thus, our conclusions are robust to numerical choices.
 

\bsp	
\label{lastpage}
\end{document}